\documentclass[pre,aps,twocolumn,showpcs,floatfix,amsmath,amssymb]{revtex4}
\usepackage{amsmath}
\usepackage{graphicx,dcolumn,bm,subfig} %natbib
\usepackage{bm}        % for math
\usepackage{mathrsfs} % for scripted letters
\usepackage{color}
\usepackage{caption}
\usepackage[utf8]{inputenc}
\usepackage{verbatim}

\usepackage{cleveref} %to reference multiple labels in one command

\def\3dots{\:\raisebox{-0.5ex}{$\stackrel{\textstyle.}{:}$}\:}

\def\beq{\begin{equation}}
\def\eeq{\end{equation}}
\def\bea{\begin{eqnarray}}
\def\eea{\end{eqnarray}}

\begin{document}
\title{Geometrical effects on mobility}
\author{Jorge H. Lopez$^1$}
\affiliation{$^1$Department of Civil Engineering, Universidad Mariana, Pasto, Colombia}
  
\begin{abstract}
In this paper we analyze the effect of randomly deleting streets of a synthetic city on the statistics of displacements. Our city is constituted 
initially by a set of streets that form a regular tessellation of the euclidean plane. Therefore we will have three types of cities, formed by squares, 
triangles or hexagons. We studied the complementary cumulative distribution function for displacements (CCDF). For the whole set of 
streets the CCDF is a stretched exponential, and as streets are deleted this function becomes a linear function and then two clear different exponentials.
This behavior is qualitatively the same for all the tessellations. Most of this functions has been reported in the literature when studying the displacements
of individuals based on cell data trajectories and GPS information. However, in the light of this work, the appearance of different functions for 
displacements CCDF can be attributed to the connectivity of the underlying street network. It is remarkably that for some proportion of streets we 
got a linear function for such function, and as far as we know this behavior has not been reported nor considered. Therefore, it is advisable to 
analyze experimental in the light of connectivity of the street network to make correlations with the present work.   
\end{abstract}

\maketitle

\section{Introduction}
There has been an increasing interest in comprehending the way cities organizes in the last few decades \cite{Batty, Makse, Bettencourt1, 
Bettencourt2}. The structure of a city influences human activities and in particular the mobility inside it \cite{Jiang1, Chen, RiccardoGalloti, 
Kang}. Some of the first works in the present century claimed that human mobility inside cities has a random nature. For example, in 
Ref.~\cite{Jiang1} it is stated that if we put randomly a set of "walkers" and let them move arbitrarily with some constraint on their time trip, the 
distribution probability for the distance traveled is the same as real human distances distribution. Amazingly then, human mobility seems not to 
depend on the purposiveness of trips but instead it relies on the particular  street network cars are moving on. This view, that human mobility is 
random in nature, has been reviewed and nowadays more focus is put  on the purpose of human trips \cite{Lee, Deville, Song, Piovani, Sun1, Sun2, 
Chodrow, Kitamura}. However, it seems that the structure of the road network is relevant in leading the mobility patterns. For example in 
~\cite{Olmos} it is shown that having blocks with a honeycomb structure is more favorable for having less traffic jams than a street network 
composed of square blocks, as many cities are built. Also, in ~\cite{Kang}, it is established that compactness and size of a city can affect human 
mobility. Therefore some attention should be put on the geometry of a street network when analyzing human mobility.

In this work we study the relationship between mobility and the underlying space through simulations. We simulate the road network and use 
different probability distributions for the speed of cars that move through the street network to analyze the results on the complementary 
cumulative distribution function (CCDF) for displacements. Moreover, we do this analysis when a number of randomly streets are deleted. It is 
worth to mention that the CCDF of a given density distribution function $f(x)$ that depends on the random variable $x$ which takes values on 
the interval $[0,\infty)$ is given by 
\begin{equation}
F(x) = 1 - \int_{0}^{x} f(t) dt
\label{CCDF}
\end{equation}
and it represents the probability that the random variable gets a value greater than $x$.   
   
In this work we analyze the structure of a street network when some streets are deleted, then we are in the domains of percolation theory. Then, 
we analyze how some probabilistic distributions for time and car speed produce some distribution for displacements, and also how this result is 
affected by the connectivity degree of the underlying space.  

\section{Model and methods}
\begin{figure}[h!]
	\centering
	\subfloat[Square Tes.]{
  		\includegraphics[width=0.17\textwidth]{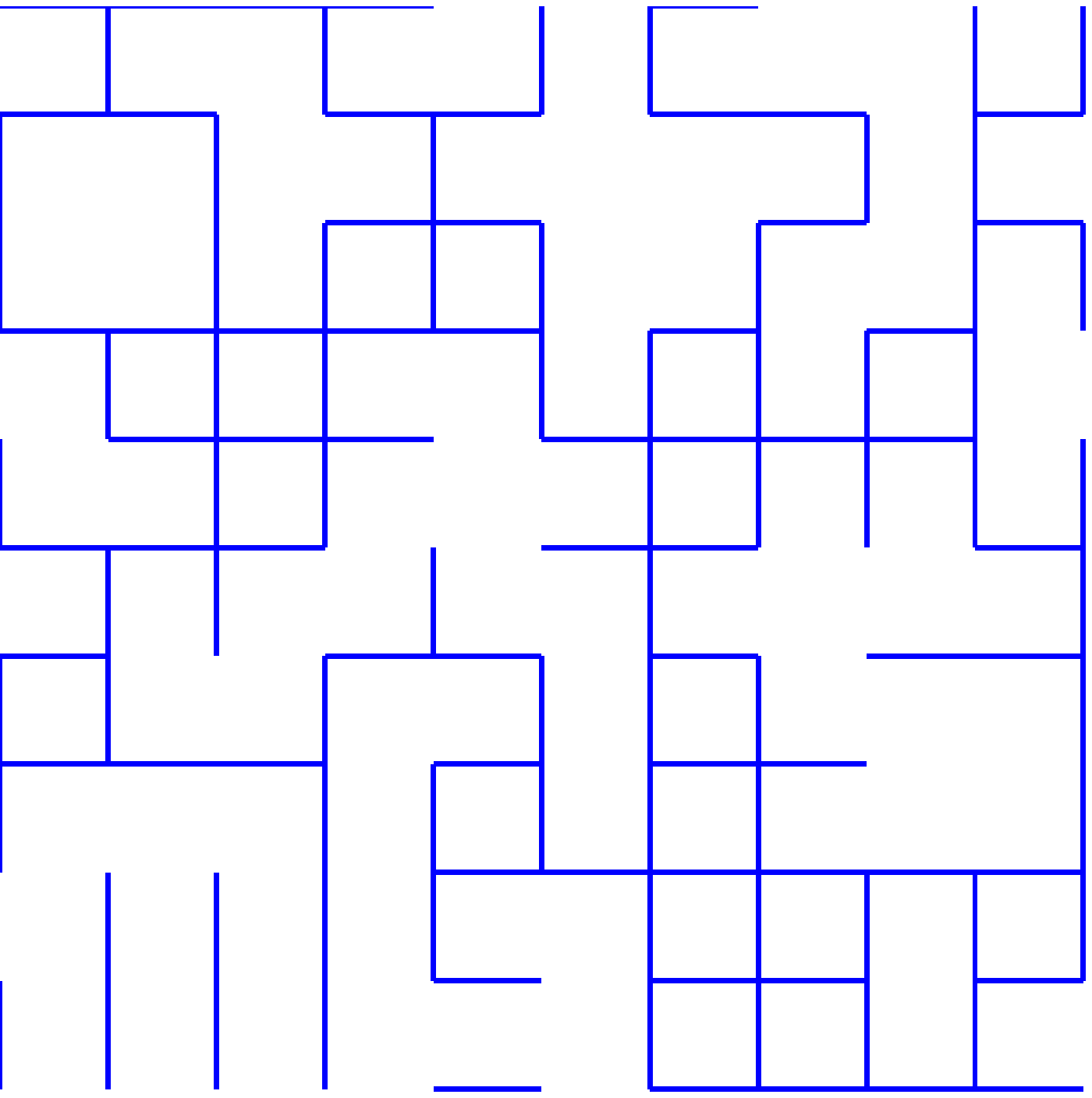}
	}
	\subfloat[Hexagon Tes.]{
  		\includegraphics[width=0.22\textwidth]{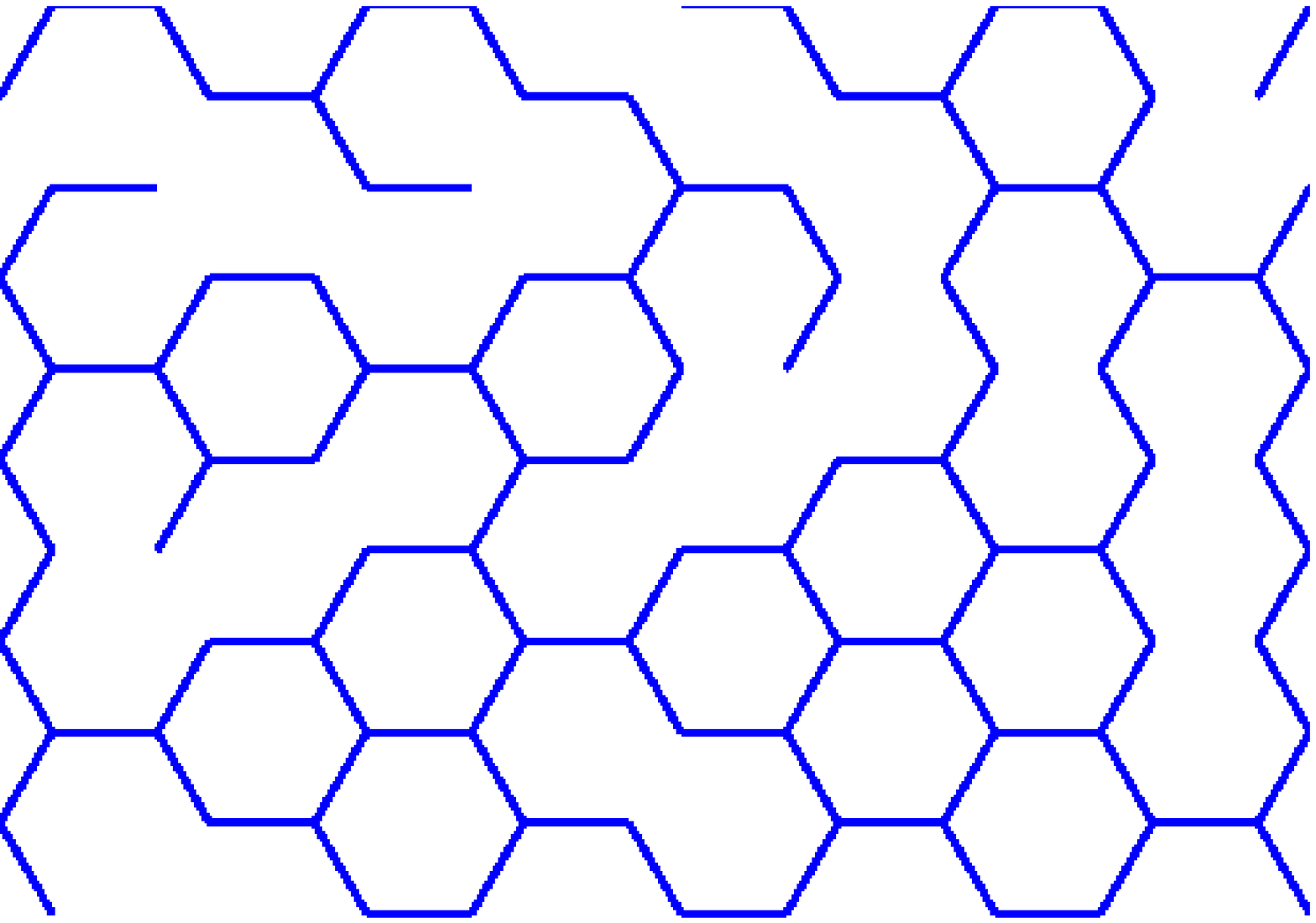}
	}
	\newline
	\subfloat[Triangle Tes.]{
  	\includegraphics[width=0.23\textwidth]{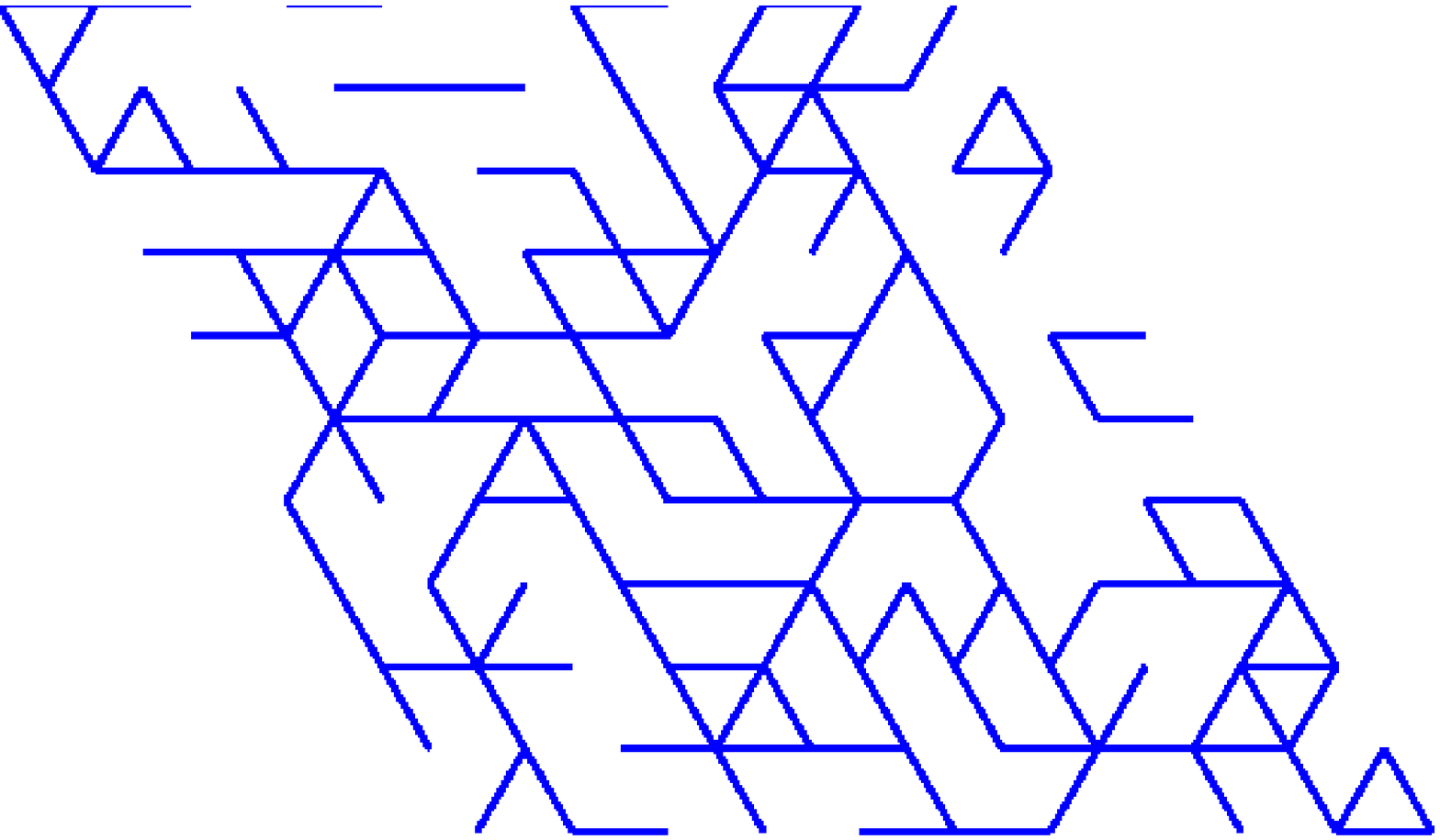}
	}
	\caption{Ilustration of synthetic cities with some streets deleted.}
	\label{Tessellations}
\end{figure}

In our simulations we start by building a road network based on some tessellation of the euclidean space. A $\{P,Q\}$ regular tessellation, 
according to the schl\"{a}fli notation, is a partition of the euclidean plane into $P$-sided regular polygons such that $Q$ polygons meet at each 
vertex. The sides and vertices of these polygons represent the blocks and corners of our street network. It is known \cite{Coxeter} that two 
dimensional euclidean space can support those tessellations that satisfy the relation
\begin{equation}
(P-2)(Q-2) = 4
\end{equation}
Accordingly, we just can tessellate the euclidean plane with triangles $(\{3,6\})$, squares $(\{4,4\})$ and hexagons $(\{6,3\})$. In this work 
we analyze each of these possibilities. Once we have a tessellation we randomly delete a portion of streets to see its implications 
on the studied variables, as illustrated on figure ~\ref{Tessellations}. 

\begin{comment}
\begin{minipage}{\linewidth}
	\centering
	\begin{minipage}{0.45\linewidth}
		\begin{figure}[H]    	
     		\includegraphics[width=\linewidth]{Square_0_6.eps}
     		\caption{Figure 1}
        		\label{Tessellations1}
        	\end{figure}
  	\end{minipage}
  	\hspace{0.05\linewidth}
  	\begin{minipage}{0.45\linewidth}
  		\begin{figure}[H]
  		\includegraphics[width=\linewidth]{Triangle_0_4.eps}
  		\caption{Figure 2}
        	\label{Tessellations2}
		\end{figure}        	
	\end{minipage}
   	\hspace{0.05\linewidth}
   	\begin{minipage}{0.45\linewidth}
   		\begin{figure}[H]
        	\includegraphics[width=\linewidth]{Hexagon_0_8.eps}
		\caption{Figure 3}        	
        	\label{Tessellations3}
        	\end{figure}
  	\end{minipage}
  	\label{Tessellations}
\end{minipage}
\end{comment}

Therefore our work relates with percolation theory. Percolation is a widely studied field in physics \cite{Stauffer, Saberi}, and the 
central problem can be state in the following way. There is a set of sites (site percolation) or bonds (bond percolation) with some connections 
between then, and there is a probability $p$, $0\leq p \leq 1$, that an element be occupied, otherwise it is empty. Then clusters are formed 
by occupied neighbor elements (sites or bonds). Then, the cluster structure variation with probability $p$ is studied. 
A major result in percolation is the existence, for an infinite space, of a particular value for $p$, called critical probability $p_{c}$, below 
which there is not a cluster that traverses or percolates the structure, and above $p_{c}$ there is a percolating cluster. The behavior of the 
clusters around $p_c$ captures some elements of a phase transition. In our case we are concerned with bond percolation as streets are 
associated with bonds. The spaces we study have critical values $p_{c}=0.347, 0.5, 0.653$ for the triangle, square and hexagonal  structures 
respectively. And we study only more realistic cases where $p>p_{c}$ to guarantee the connection of the whole city through some cluster.  

Once we have our street network, which is a percolated structure, we put randomly $N$ cars on it. Then we allow them to move with a 
given speed during some time, and each of these variables follows a predetermined probability distribution. For time we use an exponential 
distribution \cite{Galloti1, Zhao1, Galloti2, Liang}, and gaussian \cite{Pratim} distributions for speeds. The motion of cars will be constrained 
by the space in the sense it is not allowed to moved when it finds a dead-end.

\section{Results}
We study the structure of the street network and then the CCDF and probability density for car displacements.  

\subsection{Length street distribution}
\begin{figure}[h!]
	\includegraphics[width=0.35\textwidth]{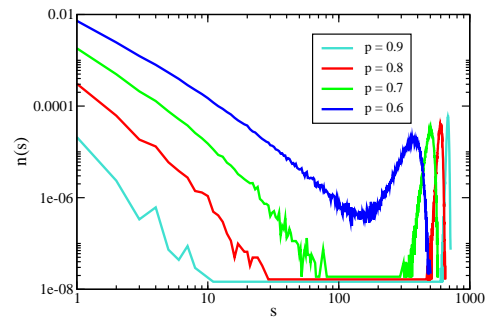} 	
	\caption{Cluster size distribution for square tessellation and $p=0.6, 0.7, 0.8, 0.9$.}
	\label{fig1:ClusSizeDistr}
\end{figure}

In order to study the street network we analyze what is called the cluster size distribution $n(s)$, that represents the average number of 
clusters of size $s$ per number of bonds (blocks). Notice that the cluster size distribution is related with the street length distribution, as the 
size of a cluster is determined by the number of blocks that it contains. It is worth to mention that the common convention in the percolation 
community is not to take into account the percolating cluster in the calculation of $n(s)$, however as we want to measure the whole structure 
of streets, we include all the clusters for calculating $n(s)$. To determine $n(s)$ we use partially the well known Newmann-Ziff algorithm 
\cite{Newman}. In Fig. \ref{fig1:ClusSizeDistr} we illustrate the behavior of $n(s)$ vs $s$ for our square city, which has a size of approximately 
20 by 20 blocks, and percolation probabilities $p=0.6, 0.7, 0.8, 0.9$.  

\vspace{3mm} 

We notice that all the curves in Fig. \ref{fig1:ClusSizeDistr} seem to behave as power laws and then appear some peak due to the finite size of
the system. The power law regime of this curves can be fitted by the expression
\begin{equation}
n(s) \sim s^{-\theta}
\end{equation}
where $\theta \in [2.0,3.0]$ and its exact value depend on $p$. We can notice also that for lower values of $p$ those curves tend to have 
a weaker tail and could be also approximated by a distribution of the form
\begin{equation}
n(s) \sim s^{-\theta}\exp\left(-Cs^{\alpha}\right)
\end{equation}

It should be mentioned that for infinite structures it is known \cite{Ding} that cluster size distribution (without a percolating cluster) satisfies 
relations
\begin{equation}
n(s) \sim \begin{cases} s^{-\theta}e^{-s/s_{\xi}}, & \mbox{$p<p_{c}$ as $\dfrac{s}{s_{\xi}} \to \infty$} \\ 
s^{-\tau}, & \mbox{$p=p_{c}$} \\ 
s^{-\theta '}e^{-\alpha s^{1/2}}, & \mbox{$p>p_{c}$}  \end{cases}
\end{equation}
Therefore the cluster size distribution is power law just at the transition probability, truncated power law below it, and what some call an 
stretched exponential above it. It is worth to mention that according to \cite{Galloti1} the displacement distribution of some vehicles follows 
also an stretched exponential.

\subsection{Displacement distribution}
In our simulation we randomly lay out $N=1000$ cars over a street network and a time given by an exponential distribution
is assigned to each car. According to \cite{Galloti1, Zhao1, Galloti2, Liang} an exponential function can model the 
time distribution for some set of vehicles. Accordingly, the time $t$ a vehicle can displace through the given space is determined by the 
function
\begin{equation}
f_{T}(t) = \dfrac{1}{\overline{t}}\exp\left(\dfrac{-t}{\overline{t}}\right)
\label{time_distribution}
\end{equation}  
where the average time is given by $\overline{t}=1800s$, following \cite{Galloti1}. When a car finds an intersection it will take another 
block randomly and it will stop when its time is finished or it finds a dead-end to analyze the effect of the network geometry on displacements. 
The speed distribution for vehicles does not seem to follow some clear pattern. However, it has been determined \cite{Pratim} that on 
highways speed vehicles tend to follow a normal distribution. Therefore, we choose the following gaussian distribution to select the speed of a 
car
\begin{equation}
f_{V}(v) = \dfrac{1}{\sqrt{2\pi}\sigma}\exp\left(-\dfrac{(v-\overline{v})^{2}}{2\sigma ^2}\right)
\label{speed_distribution}
\end{equation}
where the average speed is $\overline{v}=8m/s$ and $\sigma=0.3\overline{v}$. Our synthetic  `square city' is made of around 360 blocks 
where one block has a length of 30 units. Similar sizes we have for the two other tessellations.  Therefore the diagonal of our whole city is 
around 800 units. As the product of averages $\overline{v}\overline{t}=3600$ then most of the cars can scatter through the whole city. 
We should notice that just $v \geq 0$ speed values make sense, therefore we choose the absolute value to be assigned to a speed when 
$v<0$.

To analyze our results we use the free software grace \cite{Xmgrace} and for interpolation we check among the functions: stretched 
exponential, power law, exponential, linear, polinomial, logarithmic and inverse functions.

\begin{figure}[h!]
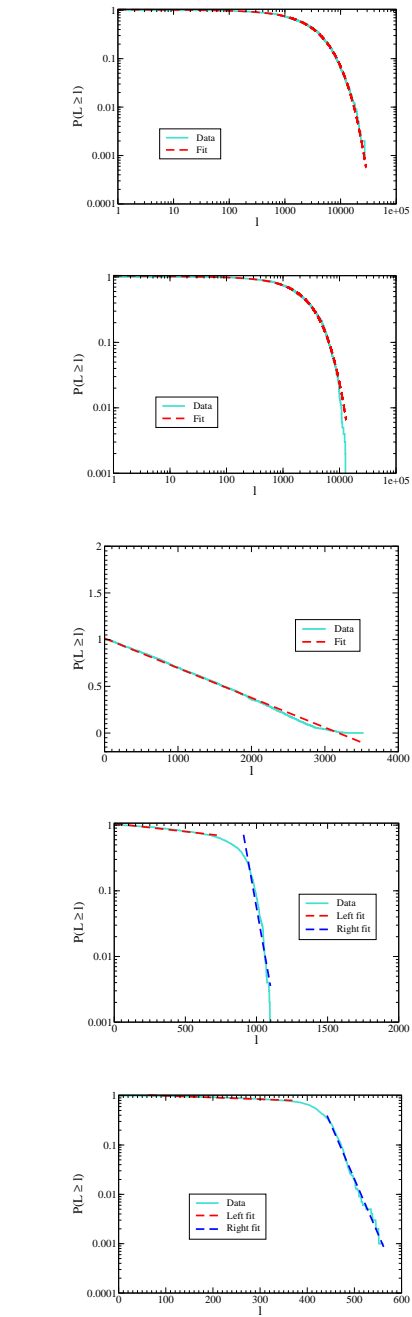

	\centering
	\subfloat[$p = 1.0$]{
  	\includegraphics[width=0.25\textwidth]{Displ_distr_Squ_Car1000_samples270_p_1_0_grace.eps}
  	\label{Square_displ_p_1_0}
  	}
	\vspace{0.1cm}  	
  	\subfloat[$p = 0.9$]{
  		\includegraphics[width=0.25\textwidth]{Displ_distr_Squ_Car1000_samples390_p_0_9_grace.eps}
  		\label{Square_displ_p_0_9}
	}
	\vspace{0.1cm}
	\subfloat[$p = 0.8$]{
  	\includegraphics[width=0.25\textwidth]{Displ_distr_Squ_Car1000_samples330_p_0_8_grace.eps}
  	\label{Square_displ_p_0_8}
	}
	\vspace{0.1cm}
	\subfloat[$p = 0.7$]{
  		\includegraphics[width=0.25\textwidth]{Displ_distr_Squ_Car1000_samples390_p_0_7_grace.eps}
  		\label{Square_displ_p_0_7}
	}
	\vspace{0.1cm}
	\subfloat[$p = 0.6$]{
  		\includegraphics[width=0.25\textwidth]{Displ_distr_Squ_Car1000_samples390_p_0_6_grace.eps}
  		\label{Square_displ_p_0_6}
	}
	\caption{Illustration of the CCDF of displacement distribution for the square tessellation and various values for $p$.}
	\label{Square_displ}
\end{figure}

The displacement CCDF results for the square tessellation are ilustrated in Fig. \ref{Square_displ} and table \ref{Table_Square} contains the 
respective regressions.  

\begin{table}[t]
\caption{Curve regressions for square city}
\centering
\label{Table_Square}
\begin{tabular}{rlc}
\noalign{\smallskip} \hline \hline \noalign{\smallskip}
$p$ & $\quad \quad \quad \quad \quad \quad$ Regression & $R^2$ \\
\hline
\vspace{2mm}
1.0 & $P(L>l) = 1.03l^{-0.0064}e^{-0.00031l^{0.98}}$ & 0.9997 \\
\vspace{2mm}
0.9 & $P(L>l) = 1.05l^{-0.0096}e^{-0.00011l^{1.13}}$ & 0.9994 \\
\vspace{2mm}
0.8 & $P(L>l) = 1.01 - 0.00032l$ & 0.987 \\
0.7 & $P_{left}(L>l) = 1.06e^{-0.00057l}$ & 0.961 \\
\vspace{2mm}
      &  $P_{right}(L>l) = 8.46\times10^{10}e^{-0.028l}$  &  0.935 \\ 
0.6 & $P_{left}(L>l) = 1.08e^{-0.00081l}$  &  0.945 \\
      & $P_{right}(L>l) = 2.49\times10^{9}e^{-0.051l}$  &  0.991 \\
\noalign{\smallskip} \hline \noalign{\smallskip}
\end{tabular}
\end{table}

For the whole structure, i.e. $p=1$, the  displacement distribution is the statistical product between the speed distribution and time distribution 
as in this case there is not any spatial constraint. It is known that, if $f_{V}$ and $f_{T}$ represents the probability density functions for speed 
and time respectively, then the probability density distribution for displacements is given by 
\begin{equation}
f_{L}(l) = \int_{0}^{\infty} f_{V}(v)f_{T}(l/v)\dfrac{1}{v} \, dv
\end{equation}
Then the respective CCDF function is 
\begin{equation}
F(l) = 1 - \int_{0}^{l} \int_{0}^{\infty} f_{V}(v)f_{T}(l/v)\dfrac{1}{v} \, dv \, dl
\end{equation}
And then we have 
\begin{equation}
\begin{split}
F(l) \, = \, & 1 - \int_{0}^{l} \int_{0}^{\infty} \dfrac{\left(\overline{l/v}\right)^{-1}}{\sqrt{2\pi}\sigma}v \\
         & \exp\left(-\dfrac{(v-\overline{v})^2}{2\sigma^2}-\dfrac{l\left(\overline{l/v}\right)^{-1}}{v}\right) \, dv \, dl
\end{split}
\end{equation}
This curve was so well aproximated by the stretched exponential 
\begin{equation}
P(L>l) = 1.03l^{-0.0064}e^{-0.00031l^{0.98}}
\end{equation}
having an square correlation coefficient $R^{2}=0.9997$. Notice that according to ~\cite{Galloti1} experimental data is also better 
approximated by an stretched function although with different constants. For $p=1.0$ we have very similar results for all the tessellations as it 
is expected from the fact that cars can move freely on the entire space. For the case $p=0.9$ (see Fig. ~\ref{Square_displ_p_0_9}), i.e. when 
10\% of the streets are deleted randomly, we have again an stretched exponential function, although with some minor variations on the 
parameters. The more drastic change occurs for $p=0.8$ as the CCDF is better approximated by a linear relationship as illustrated on figure 
~\ref{Square_displ_p_0_8}. As far as we know this type of distribution has not been reported on the literature for human mobility but 
we have a linear relation with a high correlation coefficient $R^{2}=0.987$. From the definition of a CCDF (eq. \ref{CCDF}) we have 
\begin{equation}
\dfrac{dF(x)}{dx} = -f(x)
\label{CCDF_f(x)}
\end{equation}

Therefore if a CCDF is linear then the respective probability density function is a \emph{constant}. Surprisingly then, in this case, all the 
displacements have the same probability. Somehow the interplay between the geometry (when 20\% of the streets are deleted for the square
city) and time and speed distributions make that the density probability function for displacements to be a constant. For cases 
$p=0.7, 0.6$ we have that CCDF for displacements is better approximated by two different exponentials as illustrated in Figs. 
~\ref{Square_displ_p_0_7} and ~\ref{Square_displ_p_0_6}. Regarding the transition point between the two curves we found 
that it is $l \sim 800$ for $p=0.7$ and $l \sim 400$ for $p=0.6$. Since our city is a square of length $800$ we see that for $p=0.7$ the 
transition point is comparable with the linear size of the whole city, while it is around one quarter of the city for $p=0.6$.

For `cities' with percolation probabilities $p \leq 0.5$ we got CCDF for displacements that can be exponential or power law. However for this 
type of cities the street network does not connect the whole city, and therefore this cases are not very realistic. \\

When working with the other two tessellations, hexagonal and triangular, we have the same qualitative results, i.e., stretched exponential
$\rightarrow$ linear $\rightarrow$ two exponentials when moving from $p=1.0$ to the critical value $p_{c}$ of the respective space, as 
illustrated on tables \ref{Table_Triangle} and \ref{Table_Hexagon}. Let us remember that $p_{c}=0.653$ and $0.347$ for hexagonal and 
triangle tessellations respectively. 

\begin{table}[h]
\caption{Curve regressions for triangle city}
\centering
\label{Table_Triangle}
\begin{tabular}{rlc}
\noalign{\smallskip} \hline \hline \noalign{\smallskip}
$p$ & $\quad \quad \quad \quad \quad \quad$ Regression & $R^2$ \\
\hline
\vspace{2mm}
1.0 & $P(L>l) = 1.01l^{-0.0016}e^{-0.00036l^{0.97}}$  &  0.9997 \\
\vspace{2mm}
0.9 & $P(L>l) = 1.04l^{-0.00000034}e^{-0.00057l^{0.92}}$  &  0.9996  \\
\vspace{2mm}
0.8 & $P(L>l) = 1.01l^{-0.00005}e^{-0.00020l^{1.05}}$  &  0.9995  \\
\vspace{2mm}
0.7 & $P(L>l) = 1.11l^{-0.028}e^{-0.0000055l^{1.53}}$  &  0.999  \\
\vspace{2mm}
0.6 & $P(L>l) = 1.06 - 0.00042l$  &  0.994  \\
0.5 & $P_{left}(L>l) = 1.05e^{-0.00062l}$  &  0.957 \\
\vspace{2mm}
      & $P_{right}(L>l) = 3.27\times10^{9}e^{-0.028l}$  &  0.968  \\
0.4 &  $P_{left}(L>l) = 1.09e^{-0.00096l}$  &  0.935 \\
      &  $P_{right}(L>l) = 3.24\times10^{5}e^{-0.037l}$  &  0.986 \\ 
\noalign{\smallskip} \hline \noalign{\smallskip}
\end{tabular}
\end{table}
\begin{table}[h]
\caption{Curve regressions for hexagonal city}
\centering
\label{Table_Hexagon}
\begin{tabular}{rlc}
\noalign{\smallskip} \hline \hline \noalign{\smallskip}
$p$ & $\quad \quad \quad \quad \quad \quad$ Regression & $R^2$ \\
\hline
\vspace{2mm}
1.0 & $P(L>l) = 1.01l^{-0.0015}e^{-0.00036l^{0.97}}$  &  0.9997 \\
\vspace{2mm}
0.9 & $P(L>l) = 1.09 - 0.00047l$  &  0.986  \\
0.8 & $P_{left}(L>l) = 1.14e^{-0.00086l}$  &  0.919 \\
\vspace{2mm}
      &  $P_{right}(L>l) = 5.21\times10^{10}e^{-0.041l}$  &  0.979  \\
0.7 & $P_{left}(L>l) = 1.07e^{-0.00083l}$  &  0.892 \\
      &  $P_{right}(L>l) = 9.04\times10^{8}e^{-0.076l}$  &  0.989 \\ 
\noalign{\smallskip} \hline \noalign{\smallskip}
\end{tabular}
\end{table}

Therefore, for all the three spaces, for higher values of $p$ the CCDF for displacements is an stretched exponential of the form
\begin{equation}
P(L>l) = Al^{-\gamma}\exp\left(-Cl^{\delta}\right)
\label{stretch_exponential}
\end{equation}
where $A\in[1.01,1.11]$, $C\in[5.5\times10^{-6},5.7\times10^{-4}]$, $\gamma\in[3.4\times10^{-7},2.8\times10^{-2}]$, and 
$\delta\in[0.92,1.53]$. Then it becomes a linear function
\begin{equation}
P(L>l) = A - Bl
\end{equation}
with $A\in[1.01,1.09]$ and $B\in[3.2\times10^{-4},4.7\times10^{-4}]$. And finally for lower values of $p$ it becomes more like two 
different exponentials
\begin{equation}
P_{left}(L>l) = A_{1}e^{-\beta_{1}l} \quad
P_{right}(L>l) = A_{2}e^{-\beta_{2}l}
\end{equation}
where $A_{1}\in[1.05,1.14]$, $A_{2}\in[3.24\times10^{5},8.46\times10^{10}]$, $\beta_{1}\in[5.7\times10^{-4},9.6\times10^{-4}]$, 
and $\beta_{2}\in[0.028,0.076]$. And the particular values depend on the tessellation. This change in the behavior is mainly 
attributed to the percentage of streets deleted, i.e. the value of $p$, as this simulation was performed using the same distribution for time and 
speed in all the three spaces. It was studied in \cite{Kang}, from data of eight cities in Northeast China, how some mobility variables are 
affected by the compactness and size of the city. In particular they exhibit data for CCDF displacements that seem to approximate to an 
stretched exponential or maybe two different exponentials (see Figs. 4, 6, and A.2 in \cite{Kang}) that would correlate at some degree with our 
functions. To better correlate such works with ours we would need to know the network street structure of the respective city. 

\begin{figure}[h!]
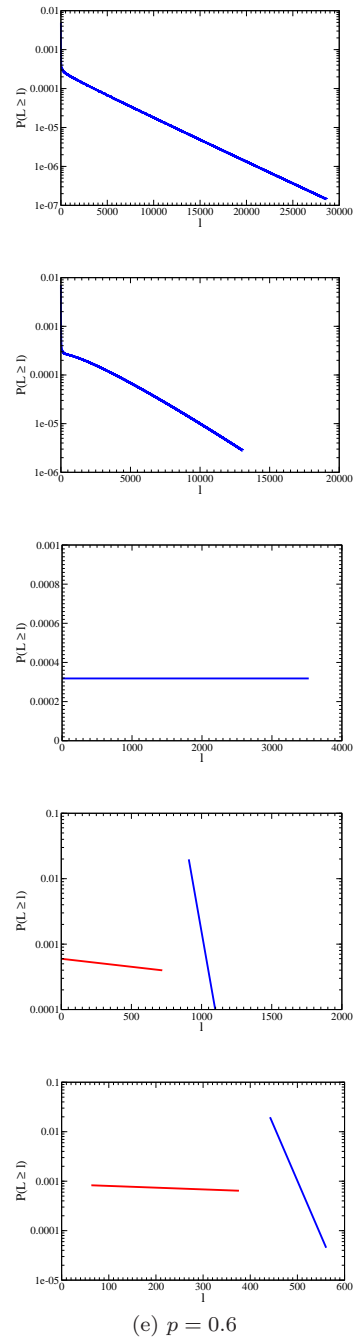

	\centering
	\subfloat[$p = 1.0$]{
  	\includegraphics[width=0.25\textwidth]{Displ_distr_Squ_Car1000_samples270_p_1_0_grace_Derivative.eps}
  	\label{Square_displ_p_1_0_f(x)}
  	}
	\vspace{0.1cm}  	
  	\subfloat[$p = 0.9$]{
  		\includegraphics[width=0.25\textwidth]{Displ_distr_Squ_Car1000_samples390_p_0_9_grace_Derivative.eps}
  		\label{Square_displ_p_0_9_f(x)}
	}
	\vspace{0.1cm}
	\subfloat[$p = 0.8$]{
  	\includegraphics[width=0.25\textwidth]{Displ_distr_Squ_Car1000_samples330_p_0_8_grace_Derivative.eps}
  	\label{Square_displ_p_0_8_f(x)}
	}
	\vspace{0.1cm}
	\subfloat[$p = 0.7$]{
  		\includegraphics[width=0.25\textwidth]{Displ_distr_Squ_Car1000_samples390_p_0_7_grace_Derivative.eps}
  		\label{Square_displ_p_0_7_f(x)}
	}
	\vspace{0.1cm}
	\subfloat[$p = 0.6$]{
  		\includegraphics[width=0.25\textwidth]{Displ_distr_Squ_Car1000_samples390_p_0_6_grace_Derivative.eps}
  		\label{Square_displ_p_0_6_f(x)}
	}
	\caption{Illustration of the probability density functions for displacement distribution on the square tessellation and 
	$p=1.0, 0.9, 0.8, 0.7, 0.6$.}
	\label{Square_displ_f(x)}
\end{figure}

We have studied CCDF because it is easier to calculate numerically compared to the probability density function because we do not need to 
handle such methods as presented in \cite{Clauset}. However, according to Eq. \ref{CCDF_f(x)} we can get a probability density function from 
the respective CCDF just by taking the negative of its derivative. In Fig. \ref{Square_displ_f(x)} we can see the probability density functions of 
the CCDFs illustrated in Fig. \ref{Square_displ}. We should mention that we did not take derivatives on CCDF data but on the regression curves 
because simulation data has many discontinuities and its respective derivative becomes very noisy. 

For $p=1.0, 0.9$ CCDF for displacements are roughly exponentials as $\gamma$ exponent in Eq. \ref{stretch_exponential} is very close to 
zero, therefore their respective probability density functions should be almost exponential as illustrated in Figs. \ref{Square_displ_p_1_0_f(x)} 
and \ref{Square_displ_p_0_9_f(x)}. In these cases, the probability density function is exponential with an exponent in the interval 
$[-0.00037,-0.00026]$. As it was mentioned previously, the case for $p=0.8$ is the most surprising as the probability density function 
becomes a constant equal to 0.00032 as illustrated in Fig. \ref{Square_displ_p_0_8_f(x)}. Then, for $p=0.7, 0.6$ in Figs. 
\ref{Square_displ_p_0_7_f(x)} and \ref{Square_displ_p_0_6_f(x)}, we have two different exponentials, as it was expected because the 
derivative of an exponential is again another exponential. For these last two cases the left exponential has an exponent in the range 
$[-0.00081,-0.00057]$, and $[-0.028,-0.051]$ for the right exponentials.

\section{Discussion}
According to our results we can claim that CCDF for displacements exhibit the same qualitatively result for all the three regular tessellations on 
the euclidean plane. Begining at $p=1.0$ and going downwards we get the same universal pattern: Streteched exponential $\rightarrow$ Linear 
$\rightarrow$ two exponentials. Therefore the degree of connectivity does not seem to affect this pattern but just the exact moments when it 
pass from one type of function to the other. As this behavior remains for the three geometries studied we infer that this could hold for a real 
network street as these tessellations are the only ones admitted by the euclidean plane. And it is certainly worthy to examine real data in the 
light of this work for the whole set of streets and to check the impact when some portions are blocked as it is the condition to move on 
different values of $p$.

In our simulations we have found different functions that have been reported on the literature. For networks working with most of the streets, 
i.e. for higher values of $p$, we got an stretched exponential. This theoretical result could be related with the work of Sagarra et al 
\cite{Sagarra}, where they seem to have this type of function for the CCDF on displacements according to their Fig. 4 (c). Regarding the 
probability density function on displacements we mainly obtain exponential functions. This is in agreement with some experimental
data as reported in \cite{Peng, Zhao2, Cai, Matsubara}. 

It is very surprising that we get a linear relation for the CCDF displacement function which in turn implies that the probability density 
function for displacements is a \emph{constant}. As far as we know this type of distribution has not been reported before in the  
literature. And it is very peculiar to have a phenomenon where the probability for trayectories \emph{do not} depend on their length.
Somehow, the relation among the geometry and the exponential time and gaussian speed distributions make the statistics of displacements
to be independent on their length. Actually, it is worthy to compare this phenomenon not just with human mobility but with other kind of physical
phenomena where statistics of displacements is not dependent on length. 

We also cheked the results when the average speed depends linearly on time, result that was obtained from data in ~\cite{Galloti1}. And
this comes from the fact that longer displacements tend to use different a different mode of transportation or streets with higher 
average speeds. In this case we had small variations in the results, except that the linear relation disappears and we have two exponential
instead in such cases. However, our work is more related with just one mode of transportation as we input the same speed and time distribution 
for the whole set of \textit{cars}. Therefore, the gaussian behavior is relevant to get the case where the probability density function is a constant.

It is worth to mention that in our model by changing just one parameter, $p$, we can move from one type of function to a very different one
in the displacement statistics. Therefore we are not restricted to a particular function and to adjust their parameters to some conditions, but the 
statistics of mobility can be transformed drastically by changing just one variable. This in turn allows us to have some measure of the impact of 
geometry on mobility.  

\vspace{4mm}
\section*{ACKNOWLEDGMENT}
J.H.L. acknowledges financial support from the Centro de Investigaciones CEI, Universidad Mariana.

\end{document}